# Automated identification and characterization method of turbulent bursting from single-point records of the velocity field

**Roni Hilel Goldshmid[1] and Dan Liberzon[1]**

[1] Faculty of Civil and Environmental Engineering, Technion, Haifa, Israel
E-mail: rhilel@technion.ac.il

## Abstract

A new automated method capable of accurately identifying bursting periods in single-point turbulent velocity field records is presented. Manual selection of the method sensitivity ($\tau^*$) and threshold ($\varepsilon_T$) are necessary for effective discrimination between burst periods and the background turbulent flow fluctuations (burst-free periods). The flow characteristic used for identification is the normalized "instantaneous" TKE dissipation rate levels, calculated using sliding window averaging. Use of the record root mean square and average values for normalization eliminates the need for definition of a physics-based flow-specific threshold. Instead, the suitable sensitivity range and the threshold parameters are selected based on preliminary examination of the velocity records. This, potentially, makes the method applicable for use across various flow fields, especially as it does not require resolving the burst-generation mechanism. The method performance is examined using a field obtained dataset of buoyancy driven turbulent boundary layer flow. Here, the selection of a two-fold ($\varepsilon_T = 2$) increase is used and the sensitivity of the method is examined. Spectral shapes of non-bursting periods show distinguished similarity to those of the Kolmogorov theory, while the bursting period spectral shapes vary significantly. Low resolution records of temperature fluctuations were observed to exhibit a significant decrease in temperature (scalar) dissipation rate during bursting periods. Based on this observation and additional processing, a statistical examination of temperature (scalar) dissipation rate is presented along with a normalization procedure. Future examination of additional scalar variations, i.e. particulate matter and/or gaseous pollutant concentrations, in connection with turbulent bursting periods can assist in further understanding of bursting generation and scalar transfer processes.

## 1. Introduction

Inaccuracies in numerical weather prediction models (NWP) and global climate models (GCM) are partially a result of insufficient real-time estimates of the total energy budget in the atmosphere, i.e. lack of accurate initial condition inputs to the models [1]. Inclusion of more complex physical parametrization of processes that occur on a sub grid scale can improve model accuracy [2,3]. For example, small scale turbulent bursting periods—bursts—remain an unresolved phenomenon that lacks proper modeling to be incorporated in numerical/forecasting models [4–6]. The CASES-99 study of the nocturnal boundary layer (BL) emphasizes the importance of intermittent turbulence modeling after observing multiple occurrences at night time [7]. The study suggests that the flow is not statistically steady, contrary to the major assumptions of the used Reynolds-Averaged form of the Navier-Stokes equations, which impedes correct representation of the stable BL and limits the capabilities of NWP and GMC [8]. Bursts are considered an additional turbulent kinetic energy (TKE) generation mechanism, and can be explained by coherent structure breakup assisting in turbulent production/maintenance of turbulence [9–13]. Often referred to as the 'bottleneck effect,' turbulence statistics that include bursting periods present a significant deviation from those corresponding canonical turbulence, appearing as a bump on the slope of the velocity fluctuation power density spectra [4,14]. Therefore, experimental characterization of this phenomenon is crucial for deriving physical parametrizations that would assist in obtaining more accurate forecasting results [15].

As the name suggests, the bursting periods are the eruption of TKE in the flow; these are characterized by a short-lived *relative* increase in the turbulent velocity fluctuation density, accompanied by elevated TKE dissipation rates, $\varepsilon$. A variety of definitions for the bursting phenomenon and the nature of burst generation mechanisms exist in the literature. In BL flows, these include "intermittent periods of considerable activity" [16], "intermittent turbulence" [17,18], "violent ejection events" [19], "lifted wall streaks" [20–23], and "a sudden loss of stability in a rising fluid and a more erratic motion ensues" [9]. Additionally, Kim *et. al.* [10] defined these periods as the "entire process which carries the flow from a relatively quiescent wall-model structure to a more random chaotic turbulent character," while Willmarth *et. al.* [23] further defined bursting as "special events when large contributions to turbulent energy and Reynolds stress occur. They (are) hypothesized to be abrupt breakages or ejections of fluid from stretching of hairpin vortices that are very intense and intermittent." These definitions are all vague on the *relativity* aspect, raising questions such as: what threshold defines a relatively considerable activity? Is there a threshold variation that should be considered for different flows? Do initial/boundary conditions affect the threshold? To address these questions, a clear definition of turbulent bursting must first be specified and a distinction between the observed phenomena and the generation mechanism must be clear.

Modern field instrumentation—including ultrasonic anemometers (sonics), radars, and LIDARS—are implemented in atmospheric BL observations to produce real time estimates of the initial conditions for prediction models. All of which have low spatiotemporal frequency response making bursts undetectable in real time acquisition. Sonics, for example, have a long acoustic fly path $O(0.1\,m)$ over which velocities are averaged, thereby lowering the frequency response to several Hz, especially at low mean velocities [24]. Several papers do report on bursting periods captured using sonics, but they use large scale intermittent turbulence phenomenon interchangeably with turbulent bursting periods, which are of much smaller scales [25–27]. An example of intermittent turbulence is the commonly observed phenomenon in stable BL, where the turbulence is constantly suppressed by the strong thermal stability and appears sporadically in time and space [28]. When the background/surrounding flow is not turbulent, capturing intermittent turbulence intervals is achievable. Reina and Mahrt [25] reported on the use of overlapping window averaging of normalized Reynolds stresses to detect such instances. The otherwise quiescent background flow made the capture and identification of intermittent turbulence possible using ultrasonic anemometers alone. However, when the background flow field is itself turbulent everything changes. The distinction of short periods of more intense velocity fluctuations *relative* to the turbulent background fluctuations, often referred to as turbulent bursts or bursting periods, is substantially more challenging. Violent ejections accompanied by bursting periods were first observed in the laboratory in BL flows using the hydrogen bubble technique [11], and later using dye injected flow visualization [19]. Hotwire anemometry in laboratory studies also captured the phenomenon [11,16,19,28–30], and only recently it was captured in the field using a novel collocated sonic-hot-film probe, combo probe [4,14].

Here, we report on a new method for detection of *short bursting periods* in a turbulent flow. These are defined as *periods where an increase in the velocity fluctuation density is observed relative to that of the background turbulence*. Bursting periods are identifiable in visual examination of the velocity component fluctuation time series; however, the literature does not offer a universally quantified measure and threshold to detect the beginning and end of such periods. Making a clear distinction between the phenomenon and the generation mechanism responsible for its creation, we offer a method for the phenomenon detection. And hence, a tool for future studies to investigate generation mechanisms by conducting analysis of various flow field characteristics, separately for bursting and burst-free time periods.

To obtain a general framework of a turbulent bursting generation mechanism, one can imagine an unbounded, homogeneous, and isotropic turbulent flow with an upstream obstacle generating vortex shedding. The generated vortices will break down at some point due to their inability to sustain their original shapes, hence creating time intervals of increased turbulent fluctuation density. More generally, any turbulence generation mechanism may cause intermittent bursting periods that may live in a turbulent, transitioning, or laminar background flow. As for BL turbulent flows, these are several bursting phenomenon generation mechanisms observed/proposed over the years:

1. The overturning of small-scale internal waves was observed to intermittently generate turbulence, i.e. turbulent bursting periods [31].

2. Shear instability caused by a low level jet that flows in a BL with strong stratification stability [17,18,28,29,32,33].

3. The Kelvin-Helmholtz instability on the interface between two distinct densities of the fluid may generate turbulent bursting periods. These separated vortices are foreign to the background flow and may appear as an intermittent increase in fluctuation density in point measurements [34,35].

4. Fluid ejections, in which low-speed streaks are rising from the boundary and oscillating, suddenly finding themselves in higher speed background with the potential for abrupt breakage causing an intermittent increase in the velocity fluctuation density. Such ejection processes may also be strengthened by buoyancy [36,37]. Identification of violent ejections near the boundary, often assisted by flow visualization, is then used to detect the bursts [19–23,38,39].

5. Bursting generation mechanism associated with passing of outer region intermittent structures was proposed by Offen and Kline [22] using space time correlation and conditioning of turbulent vs non-turbulent observed periods [10,11,40], additionally derived mathematically by Lohse and Grossmann [41]. The problems encountered with correlation derivations originated from the fact these were made using the entire turbulent time series and not the bursting events alone [30].

The variety of generation mechanisms proposed for the bursting phenomenon in the literature should be considered when attempting to characterize the events. Justifying/correlating the flow with any of the possible mechanisms is most accurate if the bursting periods are accurately identified and examined separately from the background flow.

The transient nature of bursting periods makes their identification in available velocity fluctuation time series a nontrivial and cumbersome task, and over the years the proposed techniques focused on the generation mechanism rather than on the observed bursting phenomenon itself. The identification of violent ejections has been studied extensively, but remains a challenge, especially if flow visualization is not available (such as in the case of conventional atmospheric field studies). The obtained results are not consistent between the detection methods due to the presence of background turbulence [16,19]. Bogard and Tiederman [19] compared several suggested methods including quadrant analysis (i.e. conditional sampling using QII), certain streamwise conditions, rate of change of the velocity components, and correlation with visual observations made using simultaneously sampled data and flow visualization. Pointing out a wide discrepancy between the detection methods, they conclude the Quadrant technique with its proposed threshold of Lu and Willmarth [20] yields the best results when compared to flow visualization.

To the best of our knowledge, the method proposed by Kit *et. al.* [4] is currently the only one available in the literature that is capable of directly detecting bursting periods, independent of knowledge or identification of the underlying generation mechanism. In their method, they proposed using a minimal threshold of varying TKE dissipation ($\varepsilon$) values, obtained at one-second long ensemble averages, determining if each selected ensemble of velocity fluctuation records contains bursting periods or is burst-free. The threshold was selected in a similar manner to the other methods [19,42]. When an ensemble contains values of $\varepsilon$ that are at least an order of magnitude larger than the mean value, it is claimed that the ensemble consists of at least one burst. This detection method, tagging each ensemble as containing bursts or being burst-free, is however not capable of detecting the beginning and end of respective bursting periods within the ensembles. Development of a new method for accurate identification of turbulent bursting periods, hence allowing discriminative analysis of bursts and the background flow, is of high importance to a variety of turbulence-related research fields. Beside the immediate benefit of the added ability to investigate the bursting generation mechanisms, such identification will allow a more detailed examination of turbulent flow characteristics in general, of scalar transport related phenomena, of heat transfer rates, and more. All of which are heavily influenced by the magnitude and frequency of bursting events [43]. The investigation of such phenomena can unfold novel physical models to advance the state-of-the-art weather/climate forecasting algorithms.

Here we describe a new automated bursting identification technique, developed to enable detection of turbulent bursting periods in unstably stratified upslope BL flow reported in Hilel Goldshmid and Liberzon [14] and to allow examination of turbulent bursting characteristics separately from and in comparison with the background turbulence. The method presented here relies on short term sliding window averaging and appropriate normalization of the velocity component instantaneous fluctuations. The suggested normalization of instantaneous TKE dissipation rate eliminates the need for obtaining flow-specific thresholds, hence offering the possibility for the new technique implementation across various turbulent flows. We provide an analysis of the detection method sensitivity to the user specified, data examination based, input-parameter of averaging non-dimensional window-size, $\tau^*$. The analysis is followed by data-specific optimal selection of input-parameters and a set of guidelines for their optimal selection in future studies. Results of turbulent burst detection using the method suggested here and data from Hilel Goldshmid and Liberzon [14] are also reported alongside statistical analysis of the low resolution temperature (scalar) dissipation rate in the duration of turbulent bursting periods that were separated from the background flow. Finally, a recipe is provided to enable application of the new identification method on data of various turbulent velocity fields. All calculational routines are freely available at [44].

## 2. Automated identification of bursts

An automated procedure is necessary when attempting to identify turbulent bursting periods within a turbulent background flow. However, such a task is nontrivial because the bursting period occurrence is not predictable and because the bursting period characteristics may vary depending both on the burst generation mechanism and on the background flow parameters. Recently, Kit *et. al.* [4] proposed using the TKE dissipation rate, $\varepsilon$, variations to flag pre-determined time intervals as bursting or burst-free ensembles for stable atmospheric BL flow. Averaging over one-second intervals allowed representative $\varepsilon$ values for each second in the examined ensemble to be obtained, observing a flow specific pattern of variations. The $\varepsilon$ was at least one order of magnitude higher when an interval partially consisted of a bursting period. This allowed selection of a specific threshold to distinguish between bursting and burst-free containing ensembles, however, the exact identification of the burst period starting and ending times was impossible. It was suggested that future studies should aim at identifying ensembles consisting of 100% turbulent bursting periods and comparing them with those 100% burst-free to investigate the burst related phenomena and characteristics of the flow.

Here, we suggest a different approach: implementing moving and overlapping window averaging to obtain variations of the TKE dissipation rate. This allows identification of the bursting period beginning and ending times within an examined data ensemble with sufficient accuracy, leading to an automated burst identification. To allow implementation in a variety of flows and over various characteristics of similar flows, a proper normalization of the examined averaged TKE dissipation rate is additionally required. We suggest selecting a short enough window to capture the phenomenon, as the window length is proportional to the smallest detectable burst length, and to use a sliding window with a step size corresponding to the velocity record sampling frequency to obtain the instantaneous variations of the TKE dissipation rate, $\varepsilon$. The instantaneous TKE dissipation rate is then normalized by the ratio between the ensemble averaged $\varepsilon$ and the corresponding root mean square ($rms$) value, as described in Equations (3)-(8).

Invoking Taylor's frozen turbulence hypothesis allows one to convert a single-point measurement to spatial gradients

$$\frac{\partial \psi}{\partial t} = -\left( \bar{u}\frac{\partial \psi}{\partial x} + \bar{v}\frac{\partial \psi}{\partial y} + \bar{w}\frac{\partial \psi}{\partial z} \right), \qquad (1)$$

where $\psi$ is any point measured quantity of the flow. Here, we examine variations of the velocity field TKE dissipation rate and later the temperature dissipation rate. Two distinct types of averages are used here for the sake of automated identification (Table 1); all of which are computed from the velocity fluctuation time-series. The first is a moving average, denoted hereinafter by an overbar $\bar{\psi}$, and representing a sliding window median. The second is an ensemble average, denoted hereinafter by angle brackets $\langle \psi \rangle$, and representing the ensemble median. The median was selected for averaging (of the velocity fluctuations) purposes to avoid biases resulting from outliers expected within bursting periods present in the flow. The instantaneous fluctuations relative to each of the averages are denoted as $\psi'$ when removing the moving average, and as $\psi''$ when removing the ensemble average. Finally, the $rms$ of the fluctuations is denoted as $\widehat{\psi}$ for the sliding window $rms$, and as $\{\psi\}$ for the ensemble $rms$.

|  | Sliding window | Ensemble |
|---|---|---|
| Average (mean or median) | $\bar{\psi}$ | $\langle \psi \rangle$ |
| Fluctuations | $\psi'$ | $\psi''$ |
| Root mean square (rms) | $\widehat{\psi}$ | $\{\psi\}$ |

Table 1 *Nomenclature of notations for averaging and rms calculations.*

To begin calculating the normalized instantaneous $\varepsilon$, variations of which will be used to identify the bursts, a moving-average window length, $\tau$, must be selected. The upper bound of this length should first be determined by preliminary data examination, and hence will depend on the examined flow field. A visual observation of the time series will approximate the typical turbulent bursting length, $\tau_{ATB}$. Here, it was selected to be $\tau_{ATB} = 1\ s$. The lower bound must be long enough to allow averaging over several typical relevant length scales of the turbulent flow. The length scale of choice for our data set is the Taylor microscale, $\lambda$, because it is a commonly used length scale that lies in the inertial subrange at which the eddy dissipation rate is still not dominated by viscosity. It is defined as

$$\lambda = \sqrt{\frac{\{u''\}^2}{\left\langle \left( -\frac{1}{\langle u \rangle}\frac{\partial u''}{\partial t} \right)^2 \right\rangle}}, \qquad (2)$$

where $u$ is the streamwise component [39,45]. This scale should be calculated prior to implementation of the identification method for each velocity record ensemble to be examined. The corresponding time period is then obtained by $\tau_\lambda = \lambda / \langle u \rangle$ [28] (from equation 1.4c). Finally, the averaging window length, $\tau$, is to be selected in the range of $10\tau_\lambda < \tau \leq 2\tau_{ATB}$ to allow both averaging over several periods of the Taylor scale vortices representative of the flow and to remain small enough to identify the bursting phenomenon. Finally, the non-dimensional window length, $\tau^* \equiv \tau/\tau_{ATB}$, is defined. This, in turn, provides a corresponding non-dimensional range to consider for the accepted sensitivity $10(\tau_\lambda/\tau_{ATB}) < \tau^* \leq 2$.

Once the window size is selected, the average instantaneous TKE dissipation rates, $\varepsilon$, are computed as the mean of three TKE dissipation terms:

$$\varepsilon = \frac{\varepsilon_u + \varepsilon_v + \varepsilon_w}{3}, \qquad (3)$$

where $\varepsilon_u$ is an estimate of the TKE dissipation rate using the streamwise velocity component (defined below), $\varepsilon_v$ and $\varepsilon_w$ are the TKE dissipation rate estimates using the longitudinal and transverse velocity components, respectively, and $\nu$ is the kinematic viscosity.

$$\varepsilon_u = 15 \frac{\nu}{\bar{u}^2} \overline{\left(\frac{\partial u'}{\partial t}\right)^2}, \tag{4}$$

$$\varepsilon_v = 7.5 \frac{\nu}{\bar{u}^2} \overline{\left(\frac{\partial v'}{\partial t}\right)^2}, \tag{5}$$

$$\varepsilon_w = 7.5 \frac{\nu}{\bar{u}^2} \overline{\left(\frac{\partial w'}{\partial t}\right)^2}. \tag{6}$$

The corresponding instantaneous fluctuations of the TKE dissipation rate are then defined as

$$\varepsilon'' = \varepsilon - \langle \varepsilon \rangle. \tag{7}$$

To allow comparison across various flows, these are to be normalized by the fluctuation $rms$ over the entire ensemble,

$$\varepsilon_N = \frac{\varepsilon''}{\{\varepsilon''\}}, \tag{8}$$

providing a normalization of the signal relative to the background flow. A MATLAB® script detailing the computation of $\varepsilon_N$ from velocity fluctuation time series is available in the Appendix and the complete dataset along with the MATLAB® code for reproduction of all the figures presented here are available online [44]. Variations of $\varepsilon_N$ are used to identify bursting periods along each ensemble. Time intervals during which the value of $\varepsilon_N$ exceeds a preselected threshold value ($\varepsilon_T$) are detected as bursting. The use of a threshold of $\varepsilon_N$ variations, a normalized signal inherently independent of the background flow properties, paves the way to possible generalization to bursting event detection in other flows. Its potential generalization lies in its independence of the physical properties of the flow, i.e. it does not depend on flow-specific dimensions, boundary conditions or forcing of the observed flow. Here, we examined a threshold value $\varepsilon_T = 2$ signifying a substantial (i.e. two-fold or a 200%) increase relative to that of the background turbulent flow, indicative of a bursting period presence. The threshold is somewhat arbitrary—it is very conservative and round number and any value larger than 1 would signify a deviation from the expected background turbulence variations—yet a selection of $\varepsilon_T = 1.8$ or $\varepsilon_T = 2.2$, for example, would provide similar results for the examined here data set. This threshold can, and sometimes should, be modified based on the nature of the examined flow field and especially the available experimental data, such as flow visualization of fast enough measurements of scalars. Future modification of the threshold can be achieved following the same statistical examination we provide below (i.e. Figure 5).

### 3. Experimental Dataset

In this section we demonstrate the use of $\varepsilon_N$ variations for identification of short bursting periods in data collected during a field experiment reported in Hilel Goldshmid and Liberzon [14]. An anabatic BL flow (upslope) was investigated in Nofit, a communal village in Israel, for eight consecutive days during the warm summer days of August 2015. The investigated flow developed on a moderate, 5.7°, slope on the southwestern part of the hill. The fine scales of velocity fluctuations were captured using the recently developed combo anemometer [4,46–48], composed of collocated ultrasonic anemometer (sonic) and two x-shaped double sensor hot-film (HF) probes. The calibration procedure developed by Kit *et. al.* [47] was followed. Briefly, the low pass filtering of both the carefully selected slow sonic records and of the simultaneously recorded fast HF records provided training sets for neural networks. Trained networks were then fed the original HF voltages and provided the *in-situ* calibrated 3D velocity field components. The complete experimental setup and calibration procedure are available in §3 of Hilel Goldshmid and Liberzon [14].

The data used in this study are the same 560 data blocks with a sampling frequency of 2 kHz from Hilel Goldshmid and Liberzon [14]. Each block includes four 60-second-long ensembles: the three already calibrated velocity components $u, v, w$ in the streamwise, longitudinal, and transverse directions respectively, and the sonic provided temperature $T$. A visual examination of the time series enabled Boolean tagging of each minute long ensemble as containing burst or being burst-free [14]. The period of the Taylor length scale for ensembles that were observed as containing bursts were $\tau_\lambda = 0.003 - 0.03\ s$. Leading to selection of the sliding averaging non-dimensional window length to be at least $\tau_{min}^* = 0.5$, ensuring averaging over well more than ten periods. Additionally, all approximated burst lengths ($\tau_{ATB}$) were observed to be longer than half a second by a visual examination of the data. Identification of bursting periods in the described data, using the above elaborated procedure, was performed. In the next section, we present the results obtained for various $\tau^*$ values to unfold the method sensitivity to the averaging window size selection. Due to negligible changes under the experimental conditions, the values for the kinematic viscosity and thermal diffusivity were assumed to be constant and taken to be equal to $\nu = 1.6 \times 10^{-5}\ m^2\ s^{-1}$ and $\alpha = 2.3 \times 10^{-5}\ m^2\ s^{-1}$. Discriminating the velocity field records into bursting and burst-free periods was followed by derivation of relevant turbulence statistics, comparison between bursting and non-bursting intervals and statistical analysis of temperature fluctuations.

### 4. Results

To test the stability and consistency of the results obtained using the proposed method, automatic detection of bursting periods is performed using the procedure described in §2 with varying non-dimensional window lengths, $\tau^*$. This test examines the proposed method sensitivity to the averaging window size selection. At the longest possible non-dimensional averaging window length of 60, limited by the ensemble length, the obtained results are compared with the method proposed by Kit *et. al.* [4] in §4.1. After detecting bursting periods, the fluctuations of the temperature variance dissipation rate in the duration of bursting periods were considered and compared with those of non-bursting periods in §4.2.

### 4.1 Window Size Sensitivity Analysis

To define bursting periods, the procedure and threshold described in §2 are used. Invoking a threshold of $\varepsilon_T = 2$, defines periods with $\varepsilon_N \geq 2$ as 100% *turbulent bursting* periods, periods with $\varepsilon_N \leq 1$ as 100% burst-free periods and hereinafter are referred to as *background turbulence* periods, and finally periods with $1 < \varepsilon_N < 2$ as *intermediate* periods. The latter are periods in which the TKE dissipation rate is elevated relative to that of the background flow, but is still smaller than the set bursting threshold, therefore the period is not guaranteed to be of bursting nature.

An additional condition for minimal non-dimensional bursting period length was invoked, based on the Taylor time scale values noted previously, and set to be $\tau_{min}^* = 0.5$. Moreover, consecutively detected bursting periods separated by a non-bursting period shorter than $\tau_{min}^*$ are considered as one continuous bursting period.

To examine the method sensitivity to window size selection, the automatic detection algorithm was applied to the entire dataset using a range of averaging non-dimensional window sizes, $\tau^*$. Window sizes in the range of $0.5 \leq \tau^* \leq 30.0$, were applied with $\Delta\tau^* = 0.1$, and with a coarser step size, $\Delta\tau^* = 1.0$, in a second range of $31.0 \leq \tau^* \leq 60.0$. The second range was computed to compare the results with the Kit *et. al.* [4] method in which averaging over the whole length of the ensemble (60 $s$) is applied.

Details of the first mapping of the method dependence on $\tau^*$ are presented in Figures 1-3. Parts (a) of the Figures display representative examples of the three velocity-components time series; parts (b) display the corresponding $\varepsilon_N$ variations obtained using $\tau^* = 0.5,1.0,2.0,3.0$; and parts (c) display a map corresponding to background turbulence, intermediate, and turbulent bursting periods as a function of $\tau^*$ from 0.5 to 30.0. The $\varepsilon_N$ signals presented in (b) are consistent with the general trend of regions with elevated fluctuation intensity thereby consistently marking regions with the observed phenomenon. The curves present a sharp increase in $\varepsilon_N$, indicating the ability to capture the instantaneous change in the flow properties. The maps in Figures 1-3 (c) demonstrate the reduction of accuracy of the beginning and end of bursting interval detection with an elongation of the averaging window.

For example, using averaging windows longer than about $\tau^* = 12$ in Figure 1 (c) results in inclusion of the two separate bursting events occurring at 30 $s$ and 40 $s$ in one window, as these times are included on both ends of the window. The result is a single event marked in the center of the averaging window at about 35 $s$. This is still a true detection but of a coarser and insufficient resolution. To better demonstrate this, Figure 1 (c) includes two green lines at the edges of the detected burst period on the $\tau^* = 20$ result line. When comparing this detected region to the time series, it displays that both bursting periods are indeed a part of the coarse resolution result. In Figure 2 (c) and Figure 3 (c) a shift in the detection region is observed at $\tau^* = 19$ and 7 respectively. This shift only indicates the center of the window thereby including the half of $\tau^*$ on each end of window center location and would consistently include the actual burst independent of window size. Such visual examination of the time series is made as a 'sanity check' when selecting a window size. Selection of a small enough window is hence essential for accurate capturing of the beginning and end of the phenomenon, presenting several events merging, or shifting off the detected bursting event time.

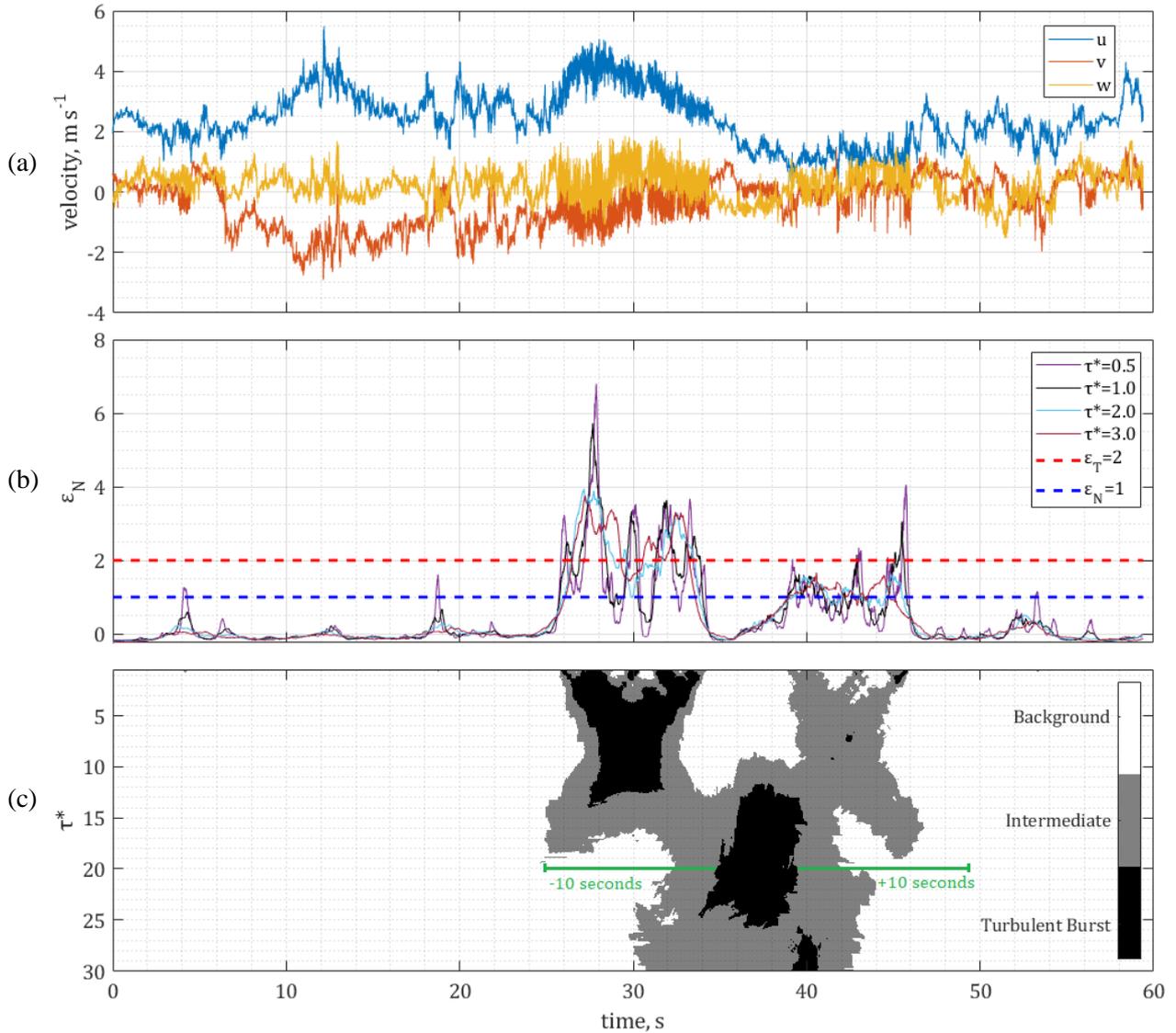

Figure 1 *(a) Time series of instantaneous velocity components $(u, v, w)$ in blue, orange, and yellow, obtained on 08-Aug-2015 at 11:56; (b) normalized instantaneous TKE dissipation rates ($\varepsilon_N$) derived using $\tau^* = 0.5, 1.0, 2.0, 3.0$ in purple, black, blue and red respectively. The red dashed line represents the maximum threshold for background turbulence periods, the area between the red dashed line and the blue dashed line represents the intermediate range, and the area above the blue dashed line represents the turbulent bursting range; (c) Displays a map of detected background turbulence, intermediate, and turbulent bursting periods in white, grey and black. These are a function of time and $\tau^*$ ranging from $0.5 - 30.0$ with $\Delta\tau^* = 0.1$. The green lines represent the size of the sensing window at $\tau^* = 20.0$. This depicts that a coarse resolution of $\tau$ can cause a merge of two separate bursting events into one due to the decreased sensitivity.*

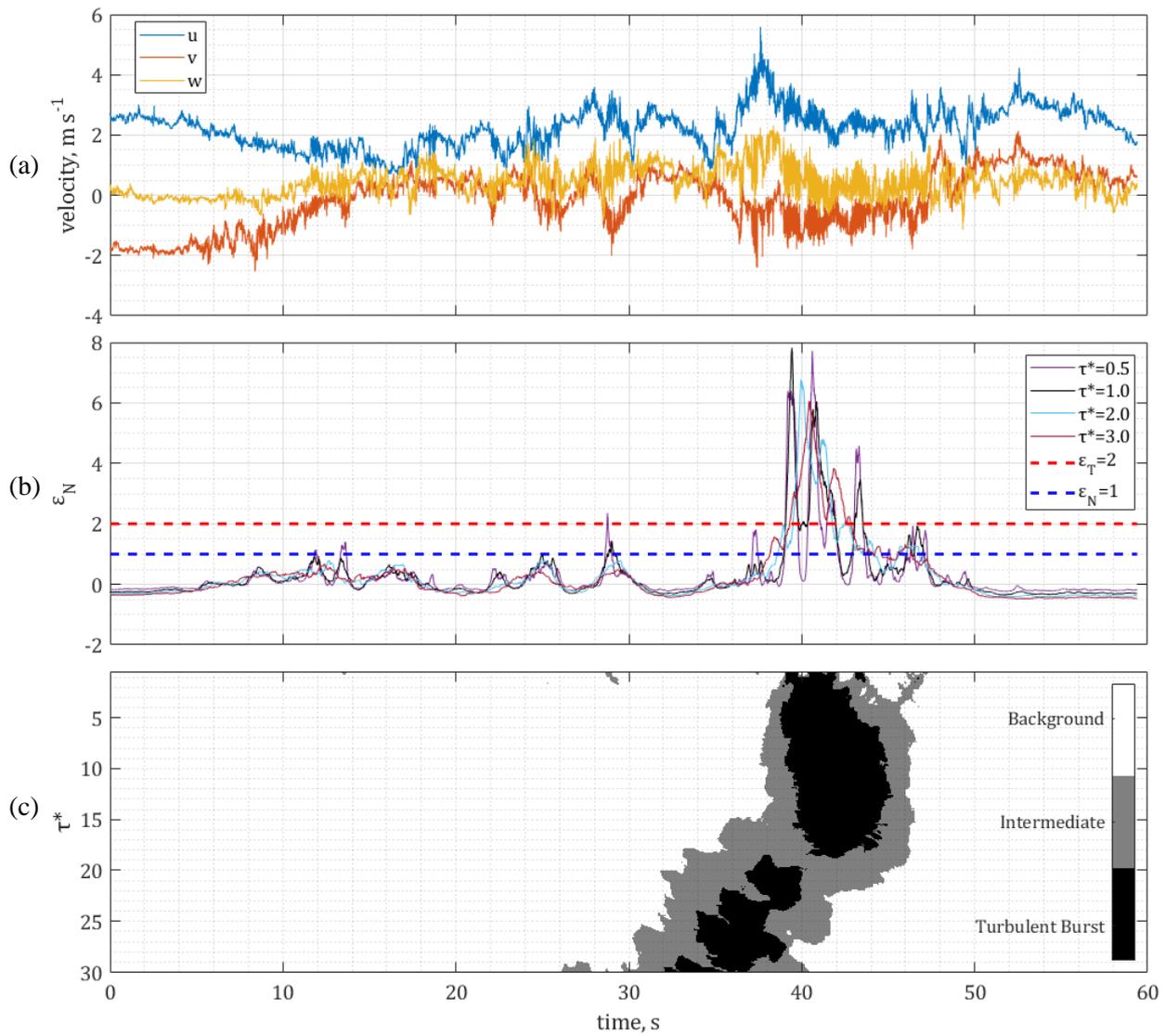

Figure 2 *(a) Time series of instantaneous velocity components $(u, v, w)$ in blue, orange, and yellow, obtained on 08-Aug-2015 at 12:28; (b) normalized instantaneous TKE dissipation rates ($\varepsilon_N$) derived $\tau^* = 0.5, 1.0, 2.0, 3.0$ in purple, black, blue and red respectively. The red dashed line represents the maximum threshold for background turbulence periods, the area between the red dashed line and the blue dashed line represents the intermediate range, and the area above the blue dashed line represents the turbulent bursting range; (c) Displays a map of detected background turbulence, intermediate, and turbulent bursting periods in white, grey and black. These are a function of time and $\tau^*$ ranging from $0.5 - 30.0$ with $\Delta\tau^* = 0.1$.*

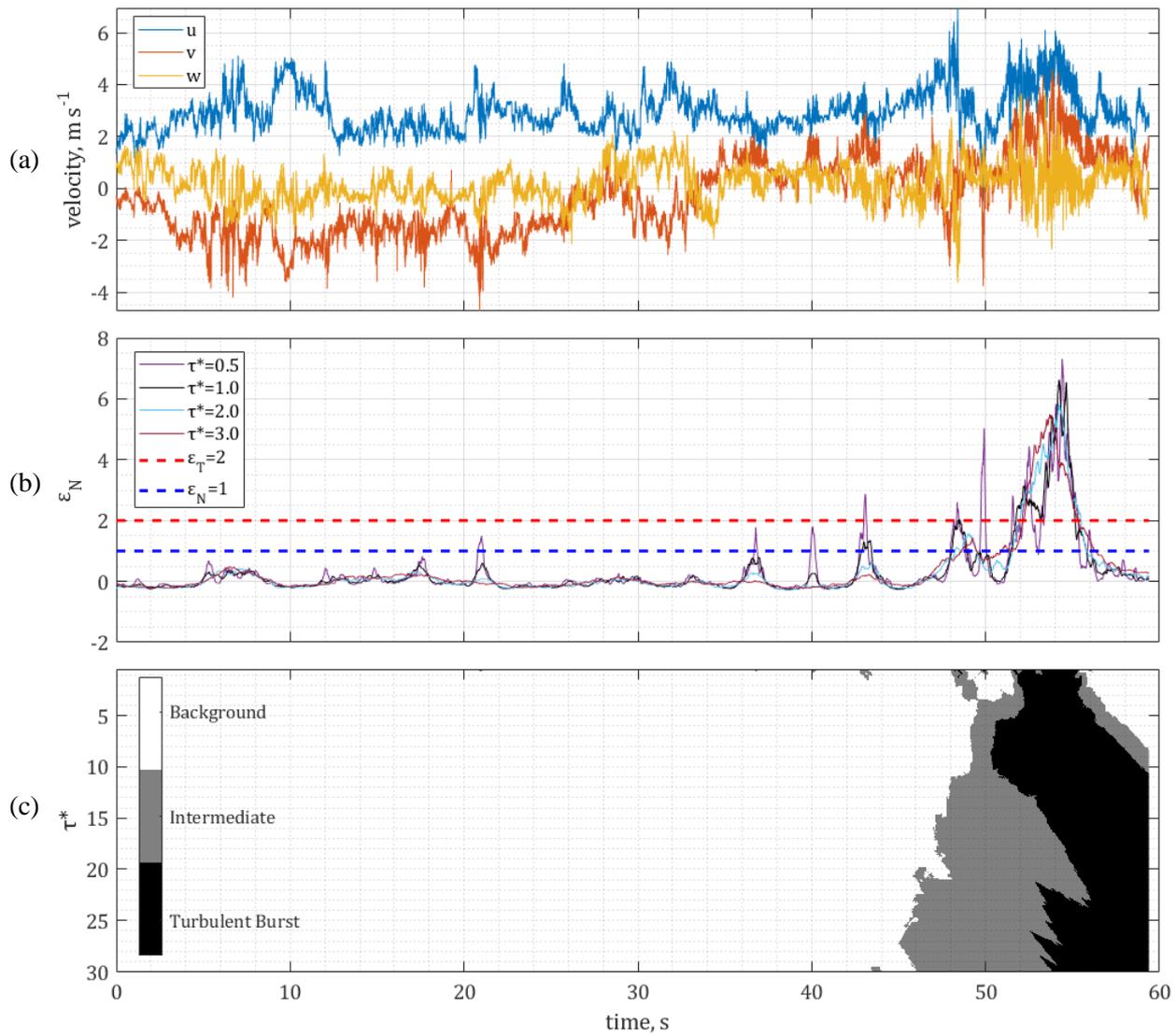

*Figure 3 (a) Time series of instantaneous velocity components $(u, v, w)$ in blue, orange, and yellow, obtained on 09-Aug-2015 at 14:20; (b) normalized instantaneous TKE dissipation rates $(\varepsilon_N)$ derived using $\tau^* = 0.5, 1.0, 2.0, 3.0$ in purple, black, blue and red respectively. The red dashed line represents the maximum threshold for background turbulence periods, the area between the red dashed line and the blue dashed line represents the intermediate range, and the area above the blue dashed line represents the turbulent bursting range; (c) Displays a map of detected background turbulence, intermediate, and turbulent bursting periods in white, grey and black. These are a function of time and $\tau^*$ ranging from $0.5 - 30.0$ with $\Delta\tau^* = 0.1$.*

After observing the general trends of the detection algorithm dependence on $\tau^*$, total counts of detected burst-periods and of one-minute long ensembles containing bursts were obtained and are displayed in Figure 4 as a function of the averaging non-dimensional window size, $\tau^*$. The ensemble count obtained for a $\tau^* = 60.0$ long window is also compared with results calculated following the Kit *et. al.* [4] method. The total burst count curve is observed to peak at $\tau^* = 1.0$, accompanied by the start of a short plateau observed in the ensembles count curve. The latter begins to decrease as the window size increases above $\tau^* = 5.0$, signifying loss of valuable instantaneous information derivable in shorter windows. The ensembles count curve reaches a minimum at $\tau^* = 40$ and the noticeable increase (which meets the Kit *et. al.* [4] values) is clear due to inclusion of the start and end of a minute effects.

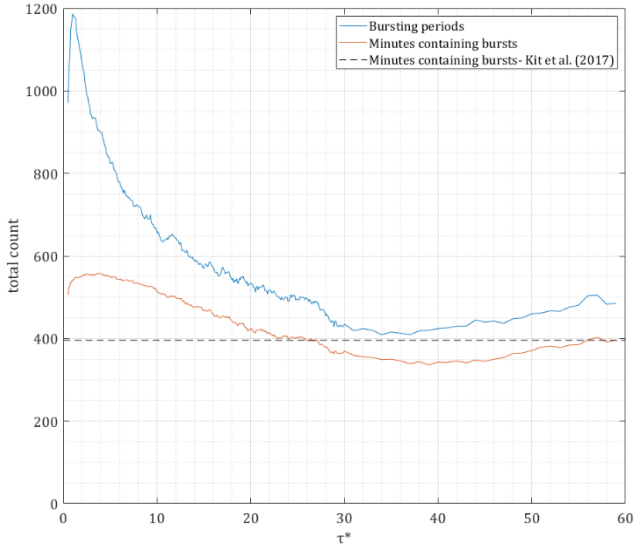

Figure 4 *Count of detected bursting periods as a function of non-dimensional window size $\tau^*$. The blue curve represents the total number of turbulent bursts detected. The orange curve displays the number of one-minute long ensembles that were tagged with at least one bursting period. The gray dashed line represents a direct comparison with the work of Kit et al [3] where one minute long ensemble average is used.*

Detecting turbulent bursting periods also allowed obtaining conditional statistics. Values of the ensemble average normalized TKE dissipation rates, $\varepsilon_N$, were calculated for each detected bursting period, thereby obtaining a representative $\langle \varepsilon_N \rangle_{TB}$ value for each bursting period. The distribution of these values is displayed in Figure 5 as probability density functions (PDF) for all examined window sizes, $\tau^* \leq 3.0$. The non-dimensional window size limit of 3.0 was arbitrary selected as large enough value, but not too large to fall into reduced accuracy detection range. The location of the peaks of all PDF distribution curves is greater than three and the areas under the curves in the range of $\langle \varepsilon_N \rangle_{TB} \geq 3$ are significantly larger than those of $\langle \varepsilon_N \rangle_{TB} \leq 3$; both observations indicate that the $\varepsilon_T = 2$ threshold is a sufficiently conservative selection. While selection of any threshold value close to 2, i.e. $\varepsilon_T = 1.8$ or $\varepsilon_T = 2.2$, would yield very similar results, which can be examined using the code [44] and modifying the threshold value. For each examined data set of turbulent velocity field fluctuations, the threshold value should be individually selected or even fine-tuned if additional types of measurements are available, e.g. flow visualization with PIV.

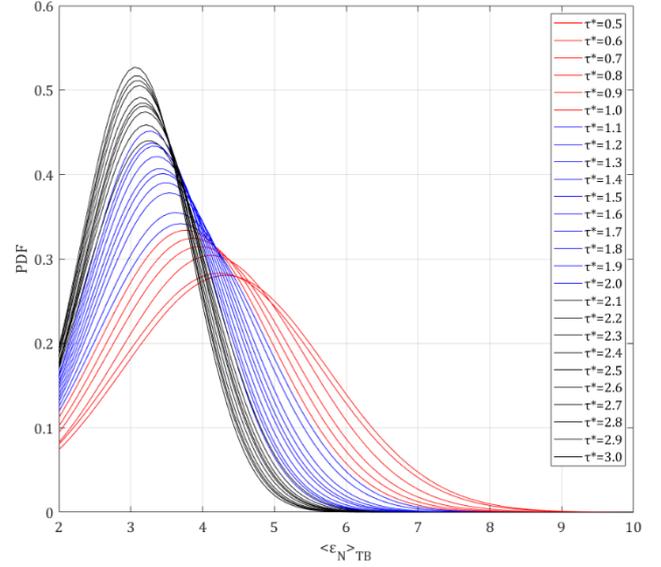

Figure 5 *Probability density functions of all detected bursting period $\langle \varepsilon_N \rangle_{TB}$ values. Each curve represents a distribution of all $\langle \varepsilon_N \rangle_{TB}$ values obtained from all detected bursts corresponding to a specific averaging non-dimensional window size $\tau^*$, while $\tau^* \leq 3.0$, as presented in the legend.*

Distribution of the non-dimensional detected burst duration, $\tau^*_{TB}$, was examined next (using a scaling similar to $\tau^*$, i.e. scaled with $\tau_{ATB}$). The use of the observation-based scaling of the method sensitivity ($\tau^*$) and of the detected turbulent bursting period lengths ($\tau^*_{TB}$) simplifies the interpretation of the obtained results and enables scalability for future studies. The exceedance probability $EP$, is an integral measure of the PDF describing the probability of an occurrence of a given value or higher. All $\tau^*_{TB}$ were ranked and the $EP$, with 1 being the largest possible value, was calculated by

$$EP = \frac{m}{n+1} \quad , \tag{9}$$

where $m$ represents the rank and $n$ represents the total number of $\tau^*_{TB}$ events [49]. Figure 6 displays the exceedance probability of $\tau^*_{TB}$, where again, different curves represent different window sizes, $\tau^*$. A decrease in an order of magnitude in the exceedance probability signifies that essentially all values beyond that range are not representative but rather are extreme cases. It is observed that for all $\tau^*$, the range of non-dimensional detected burst lengths is indeed greater than 0.5 ($\tau_{TB}^* \geq \tau_{min}^*$ as conditioned), while the maximum length of most bursting periods is between $1.5 \leq \tau_{TB}^* \leq 2.25$, just as anticipated using the visual observations and the Taylor time scales based section of the averaging window size. The short duration of the events is expected as these are transient events, not representing the background turbulent fluctuations, but representing a short-lived increase in kinetic energy due to momentum injection from some source.

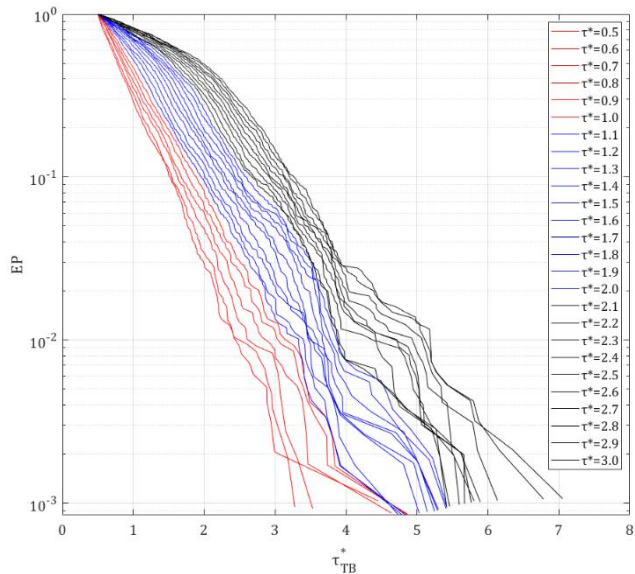

Figure 6 *Exceedance probability curves of all non-dimensional detected bursting period lengths, $\tau_{TB}^*$. Each curve represents a specific averaging window size $\tau^*$ examined, as presented in the legend.*

An additional confirmation of successful identification of bursting periods, and the now acquired ability to examine the turbulent flow characteristics separately for bursting periods and for burst-free background flow, is provided by the means of conditional spectral analysis. When bursting periods are part of a turbulent background flow, the bottleneck effect is observed in the power density spectra of velocity fluctuations [4,14]. The velocity fluctuation power density spectra of all background turbulence, intermediate, and turbulent bursting periods were calculated separately using Fourier Transforms; window averaging was implemented using one-second long windows resulting in $1\ Hz$ frequency resolution. This was performed separately for each ensemble and examination of the obtained spectral shapes showed similar findings, hence averaging for each of the three defined periods was implemented. The burst-free background turbulence period spectral shapes closely follow the Kolmogorov $-5/3$ slope in the inertial subrange. The other two cases deviate from the $-5/3$; the *turbulent bursting* periods having a notably milder slope and the *intermediate* periods have a slope in between. Figures 7-10 show the aforementioned spectral shapes obtained with $\tau^* = 0.5, 1.0, 2.0, 3.0$, respectively. The spectral shapes show that the bursting periods are characterized by higher power density of velocity fluctuations at all scales, exhibiting the smallest increase relative to the background turbulence at the larger scales, and deviating more significantly at the smaller scales. This observation corresponds to the observations of the time series plots (Figure 1-3) where the velocity fluctuation intensity during the bursts appears elevated across various scales. However, the magnitude of the elevated fluctuations is much higher at higher frequency scales. The spectral shape of the intermediate range computed with $\tau^* = 0.5$ (Figure 7) did not converge due to the small total number of intermediate range periods at the spectral resolution of $1\ Hz$.

Next, values of the Taylor microscale distribution for the various period types are displayed in Figure 11 (a)-(d). The average Taylor scales of the turbulent bursting periods, $\lambda_{TB}$, were observed to be smaller than those of the background flow turbulence $\lambda_B$, while those of the intermediate periods, $\lambda_I$, were in between. These representative average values are presented in Figure 11; they were obtained from fitting the data with exponentially modified normal distributions. This indicates the bursting periods were able to reach even smaller scales before viscous dissipation interfered, and this observation is consistently independent of the $\tau^*$ selection. The second statistical parameter observed in Figure 11 (e)-(h) is the streamwise velocity derivative skewness $Sk$, using $\tau^* = 0.5, 1.0, 2.0, 3.0$. Bursting-period $Sk$ ($Sk_{TB}$) exhibit a scatter centred around values closer to zero, indicating these values deviate significantly from the value of -0.4 observed in wind tunnel studies of grid generated homogeneous turbulence [50–53]. This behaviour is consistent regardless of $\tau^*$, indicating it is representative of the bursting events. The background turbulence period $Sk$ values, ($Sk_B$), are scattered more densely near -0.4, closely resembling the aforementioned value from wind-tunnel studies of homogeneous turbulence [50–53]. The deviation of the bursting periods from this value may indicate a deviation from the Kolmogorov theory and from the assumption that energy is supplied solely at larger scales [54–56]. This can possibly indicate that energy may also enter the system at smaller scales, or at specific bursting length scale. This claim should be examined further, in view of possible bursts generation mechanism specific for the examined flow.

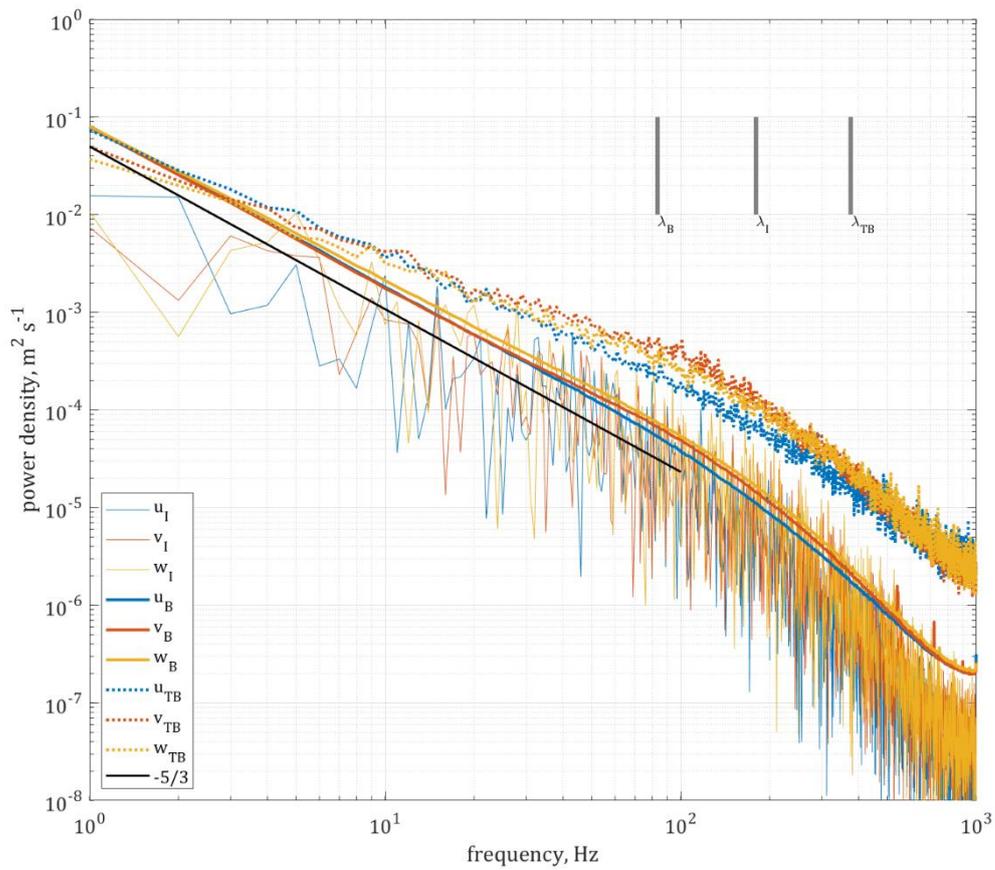

*Figure 7* Mean power density spectra of velocity fluctuations of all minute-long ensembles. The colours blue, orange, and yellow represent $u, v, w$ respectively. The curves types represent different period types in the ensembles obtained with $\tau^* = 0.5$: thick, thin, and dotted curves represent background turbulence, intermediate, and turbulent bursting. The grey vertical lines represent the Taylor scale corresponding mean frequency for background turbulence ($\lambda_B$), intermediate($\lambda_I$), and turbulent bursting ($\lambda_{TB}$) periods.

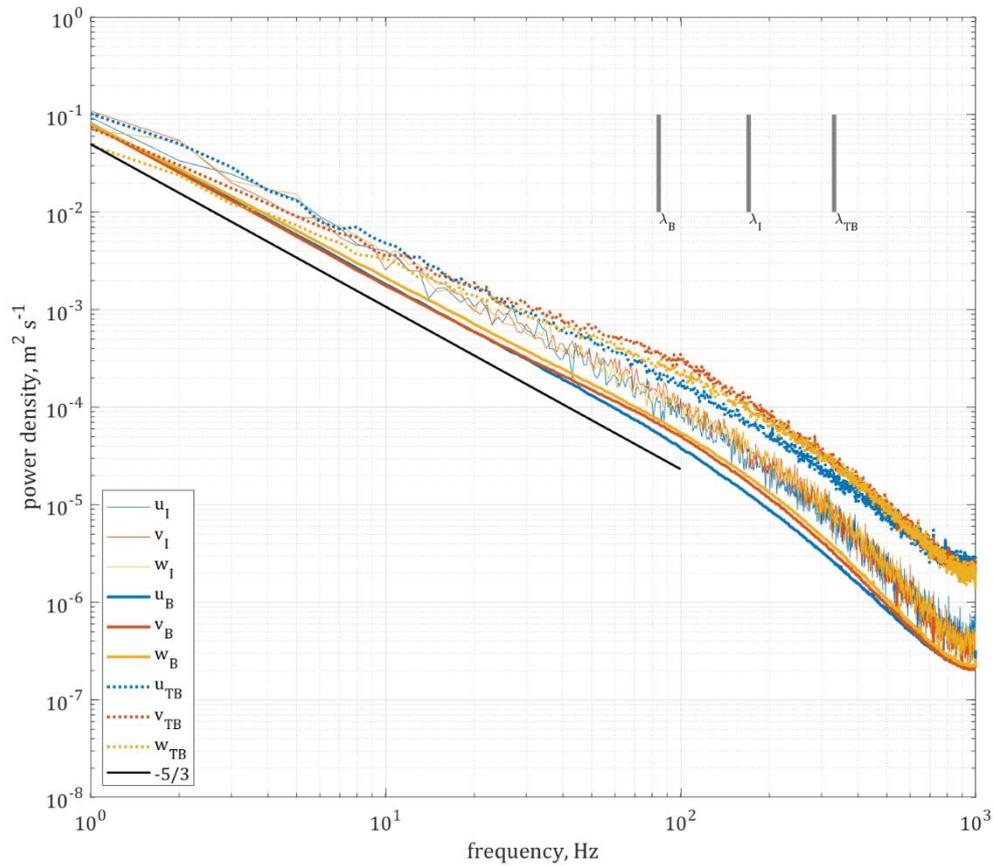

*Figure 8* Mean power density spectra of velocity fluctuations of all minute-long ensembles. The colours blue, orange, and yellow represent $u, v, w$ respectively. The curves types represent different period types in the ensembles obtained with $\tau^* = 1.0$: thick, thin, and dotted curves represent background turbulence, intermediate, and turbulent bursting. The grey vertical lines represent the Taylor scale corresponding mean frequency for background turbulence ($\lambda_B$), intermediate($\lambda_I$), and turbulent bursting ($\lambda_{TB}$) periods.

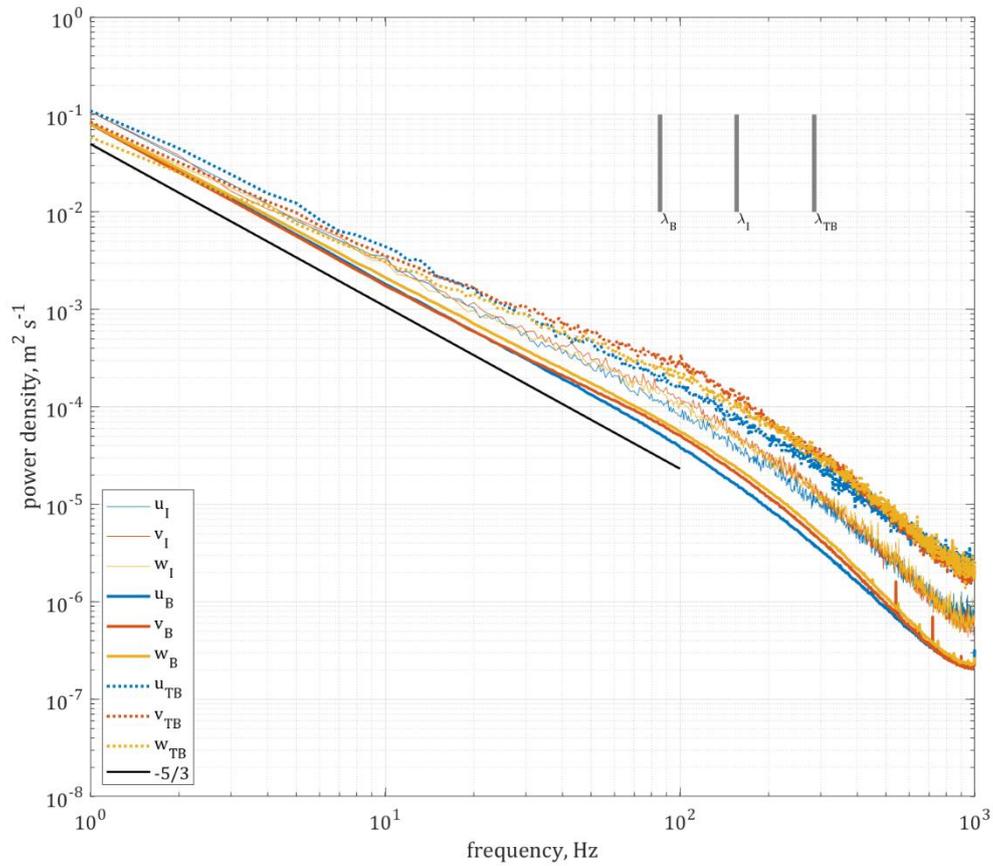

*Figure 9* Mean power density spectra of velocity fluctuations of all minute-long ensembles. The colours blue, orange, and yellow represent $u, v, w$ respectively. The curves types represent different period types in the ensembles obtained with $\tau^* = 2.0$: thick, thin, and dotted curves represent background turbulence, intermediate, and turbulent bursting. The grey vertical lines represent the Taylor scale corresponding mean frequency for background turbulence ($\lambda_B$), intermediate($\lambda_I$), and turbulent bursting ($\lambda_{TB}$) periods.

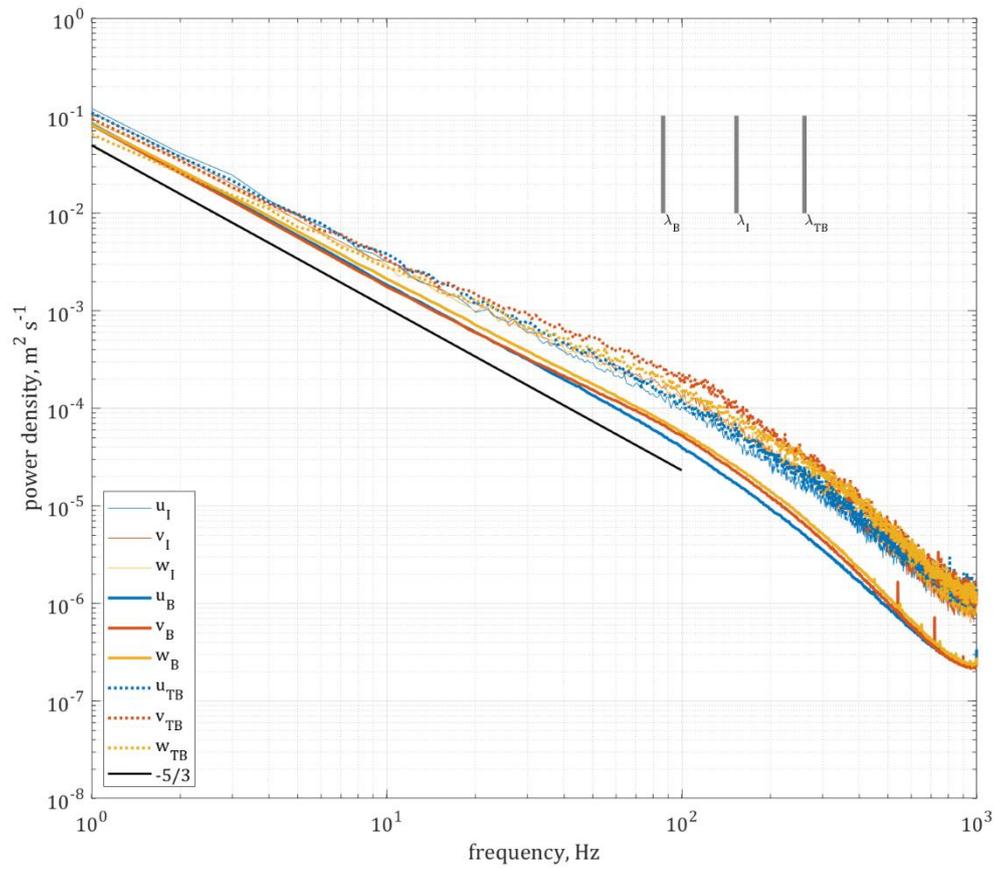

*Figure 10 Mean power density spectra of velocity fluctuations of all minute-long ensembles. The colours blue, orange, and yellow represent $u, v, w$ respectively. The curves types represent different period types in the ensembles obtained with $\tau^* = 3.0$: thick, thin, and dotted curves represent background turbulence, intermediate, and turbulent bursting. The grey vertical lines represent the Taylor scale corresponding mean frequency for background turbulence $(\lambda_B)$, intermediate$(\lambda_I)$, and turbulent bursting $(\lambda_{TB})$ periods.*

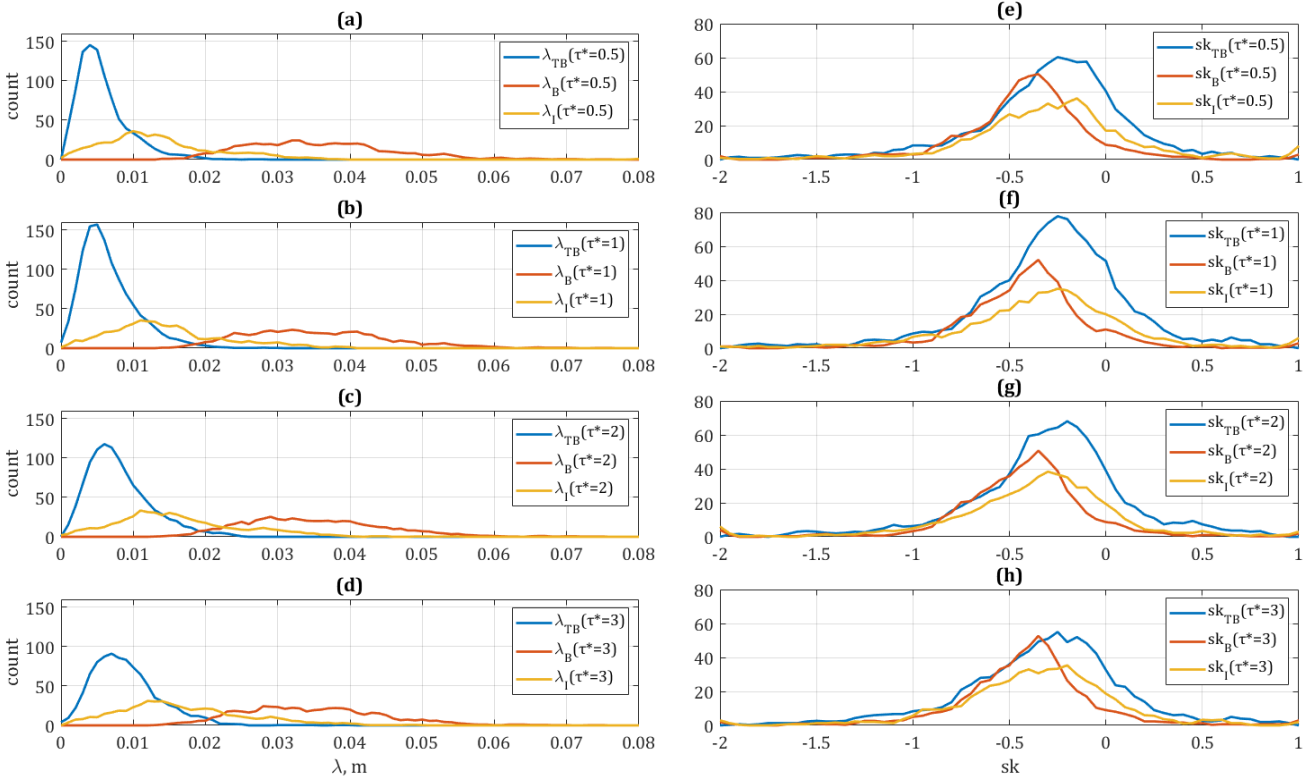

*Figure 11* Distribution of Taylor microscale, $\lambda$, and velocity derivative skewness, $Sk$, in the streamwise direction are presented at various $\tau^*$ values. The distinction between periods was made using $\tau^* = 0.5, 1.0, 2.0, 3.0$ for background turbulence, intermediate, and turbulent bursting periods, respectively. These distributions were additionally fit with an exponentially modified normal distribution to obtain the following representative peaks:

*(a)* $\tau^* = 0.5$ with peaks corresponding to $\lambda_B = 0.036$, $\lambda_I = 0.014$, $\lambda_{TB} = 0.0053$ m;

*(b)* $\tau^* = 1.0$ with peaks corresponding to $\lambda_B = 0.035$, $\lambda_I = 0.015$, $\lambda_{TB} = 0.0066$ m;

*(c)* $\tau^* = 2.0$ with peaks corresponding to $\lambda_B = 0.035$, $\lambda_I = 0.016$, $\lambda_{TB} = 0.0077$ m;

*(d)* $\tau^* = 3.0$ with peaks corresponding to $\lambda_B = 0.035$, $\lambda_I = 0.017$, $\lambda_{TB} = 0.0087$ m;

*(e)* $\tau^* = 0.5$ with peaks corresponding to $Sk_B = -0.38$, $Sk_I = -0.29$, $Sk_{TB} = -0.24$;

*(f)* $\tau^* = 1.0$ with peaks corresponding to $Sk_B = -0.40$, $Sk_I = -0.29$, $Sk_{TB} = -0.27$;

*(g)* $\tau^* = 2.0$ with peaks corresponding to $Sk_B = -0.41$, $Sk_I = -0.31$, $Sk_{TB} = -0.29$;

*(h)* $\tau^* = 3.0$ with peaks corresponding to $Sk_B = -0.39$, $Sk_I = -0.32$, $Sk_{TB} = -0.31$;

To conclude the window size sensitivity analysis, we recommend the use of the statistical and conditional spectral analysis presented above for confirmation of the threshold value $\varepsilon_T$ selection and the non-dimensional averaging window length $\tau^*$. The parameters selected here, while shown to be justified for the examined data set, most certainly will need an adjustment for use with different turbulent flow fields. Depending on various turbulent characteristics of the flow and the relevant burst generation mechanism.

### 4.2 Statistics of Temperature Observations in the Field

The ability to accurately detect periods of turbulent bursts in the velocity field measurements naturally calls for investigation of burst generation mechanism, preferably incorporating a flow visualization technique. Such an investigation can also be supported by examination of the various flow field scalar dissipation rates—such as the temperature, pressure, water vapor or some contaminant concentration, depending on the nature of a flow. Producing correlations between the scalar dissipation rates and those of the turbulent bursting and background velocity field requires obtaining the records of the relevant scalar fluctuations at resolution similar to that of the velocity field. Here we explore the temperature dissipation rate statistics aligned with detected bursting periods. While measured at much lower spatiotemporal resolution by the sonic, the observed trends of the temperature dissipation rates serve as an example of scalar investigation in view of bursts and demonstrate the technique for proper normalization.

Using the same procedure and thresholds as in §4.1, the *turbulent bursting*, *intermediate and background turbulence* periods were identified in all available ensembles and examined separately. Of interest, while examining a buoyancy-driven BL flow, were the possible changes in the temperature ($T$) behavior across the turbulent bursting periods. Changes in $T$ gradient variance are expected due to increased mixing during bursting events. And indeed, a significant decrease in $T$ variations was first noted by visual examination of the time series in the duration of a bursting period indicating uniform mixing.

Eradication of the temperature gradient variance in the duration of bursting periods (i.e. the mixing of the fluid) can be quantified using the scalar (temperature in our case) dissipation rate ($\theta$) along with proper normalization, for comparison with the normalized TKE dissipation rate variations. Each sonic-temperature time series of one-minute long ensemble, was originally sampled at $2\ kHz$ to correspond with the sampling frequency of the HF records and ensure synchronization between the two instruments recordings. Here, we down sampled the sonic provided temperature to $32\ Hz$, its declared frequency response, using sliding window averages. The instantaneous temperature dissipation rate is calculated as [57]:

$$\theta = \frac{3\alpha}{\overline{u}^2}\overline{\left(\frac{\partial T'}{\partial t}\right)^2},\qquad(10)$$

where $\alpha$ is the thermal diffusivity of the fluid.

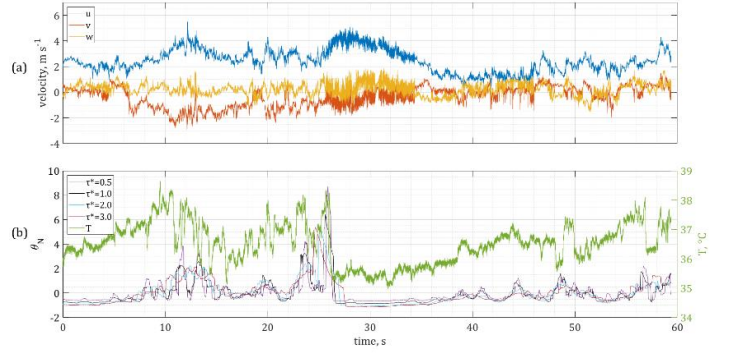

**Figure 12** *(a) Time series of instantaneous velocity components $(u, v, w)$ in blue, orange, and yellow, obtained on 09-Aug-2015 at 14:20; (b) The left axis displays $\theta_N$ as obtained using $\tau^* = 0.5, 1.0, 2.0, 3.0$ in purple, black, blue and red, respectively. The right axis displays the original oversampled temperature signal obtained by the sonic.*

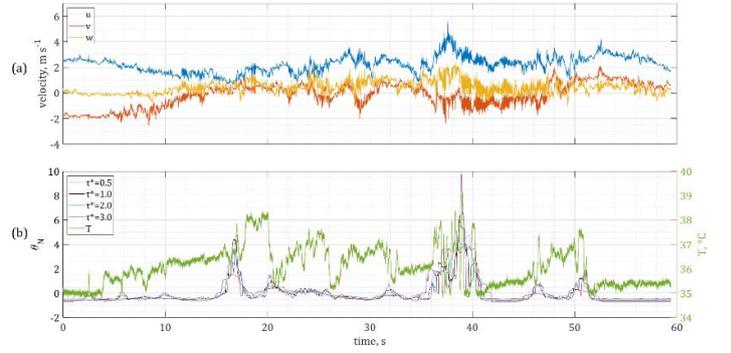

**Figure 13** *(a) Time series of instantaneous velocity components $(u, v, w)$ in blue, orange, and yellow, obtained on 08-Aug-2015 at 12:28; (b) The left axis displays $\theta_N$ as obtained using $\tau^* = 0.5, 1.0, 2.0, 3.0$ in purple, black, blue and red, respectively. The right axis displays the original oversampled temperature signal obtained by the sonic.*

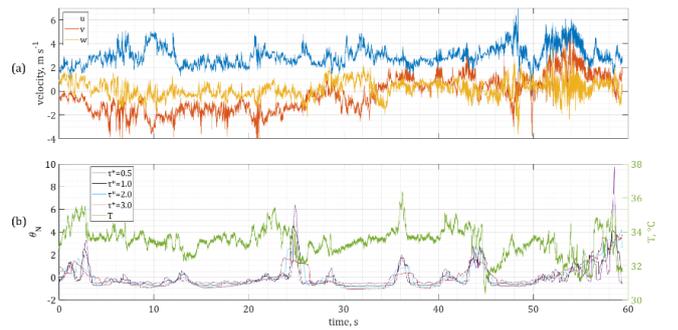

**Figure 14** *(a) Time series of instantaneous velocity components $(u, v, w)$ in blue, orange, and yellow, obtained on 09-Aug-2015 at 14:20; (b) The left axis displays $\theta_N$ as obtained using $\tau^* = 0.5, 1.0, 2.0, 3.0$ in purple, black, blue and red, respectively. The right axis displays the original oversampled temperature signal obtained by the sonic.*

Next, these instantaneous temperature dissipation rates were normalized similarly to the instantaneous TKE dissipation rate fluctuations, producing variations of

$$\theta_N = \theta'' / \{\theta''\}, \qquad (11)$$

while,

$$\theta'' = \theta - \langle\theta\rangle. \qquad (12)$$

A MATLAB® script detailing the computation of $\theta_N$ is also provided in the Appendix.

Unlike in the case of previously described TKE dissipation rate fluctuations, here the mean was used for calculation of averages. As per the sonic-limited frequency response, outliers were not expected here, any observed statistical variations may only be of large scales. The spatial derivatives from [57] are converted to temporal derivatives using the Taylor hypothesis resulting in Equation (10), with smaller values indicating the scalar is well mixed. The gradients are squared yielding only positive values that are to be averaged over time. In Equation (12), the mean of the signal is removed and in Equation (11) these new 'instantaneous variations' of the temperature dissipation rate are normalized using their respective $rms$. Therefore, negative $\theta_N$ values indicate periods during which the temperature mixing is at higher level compared to that of the background turbulence.

During (or immediately following) bursting events, a decay of temperature dissipation rate, $\theta$, is expected. Variations of $\theta_N$ derived using $\tau^* = 0.5, 1.0, 2.0, 3.0$ are presented in Figures 12-14 along with the corresponding time series of the temperature fluctuations obtained using the sonic. These Figures correspond to the same representative minutes displayed in Figures 1-3 in §4.1. Temperature dissipation rates during bursting events were found to exhibit a significantly different behaviour from no-burst periods and the trend of the temperature gradient variance suppression was investigated further.

To examine the correlation between the appearance of bursting periods and the decrease in $\theta_N$, the following processing was implemented on the entire data set. An ensemble average value $\langle\theta_N\rangle_{TB}$ representing each bursting period was derived separately for $\tau^* = 0.5, 1.0, 2.0, 3.0$; while bursting period identification for each was performed using the same values of $\tau^*$. Figure 15 presents the histograms of $\langle\theta_N\rangle_{TB}$ values for the four $\tau^*$ values. The distributions were fit with an exponentially modified normal distribution, obtaining statistical parameters representative of the examined periods. The fit-obtained averages, $\mu$, ranging between $-0.61$ to $-0.80$ and standard deviation of $\sigma = 0.1 - 0.23$, i.e. most $\langle\theta_N\rangle_{TB} \leq -0.4$. The results indicated the temperature dissipation rate experiences a significant decrease during most bursting periods, indicating higher level of scalar mixing and supporting the observations made by visually examining the time series.

The suppression of the temperature dissipation rate relative to those characterizing the background turbulent flow show a significant statistical variation, but—given the limited frequency response of the sonic—no physical interpretations can be made based on this data set beyond the demonstration of the ability to detect such variations using the new method. Taking these findings into consideration, for future studies it is recommended to perform a series of higher spatiotemporal resolution observations of a scalar of interest, e.g. using cold wires for temperature measurements, along with flow visualization techniques to better understand the physics of the flow.

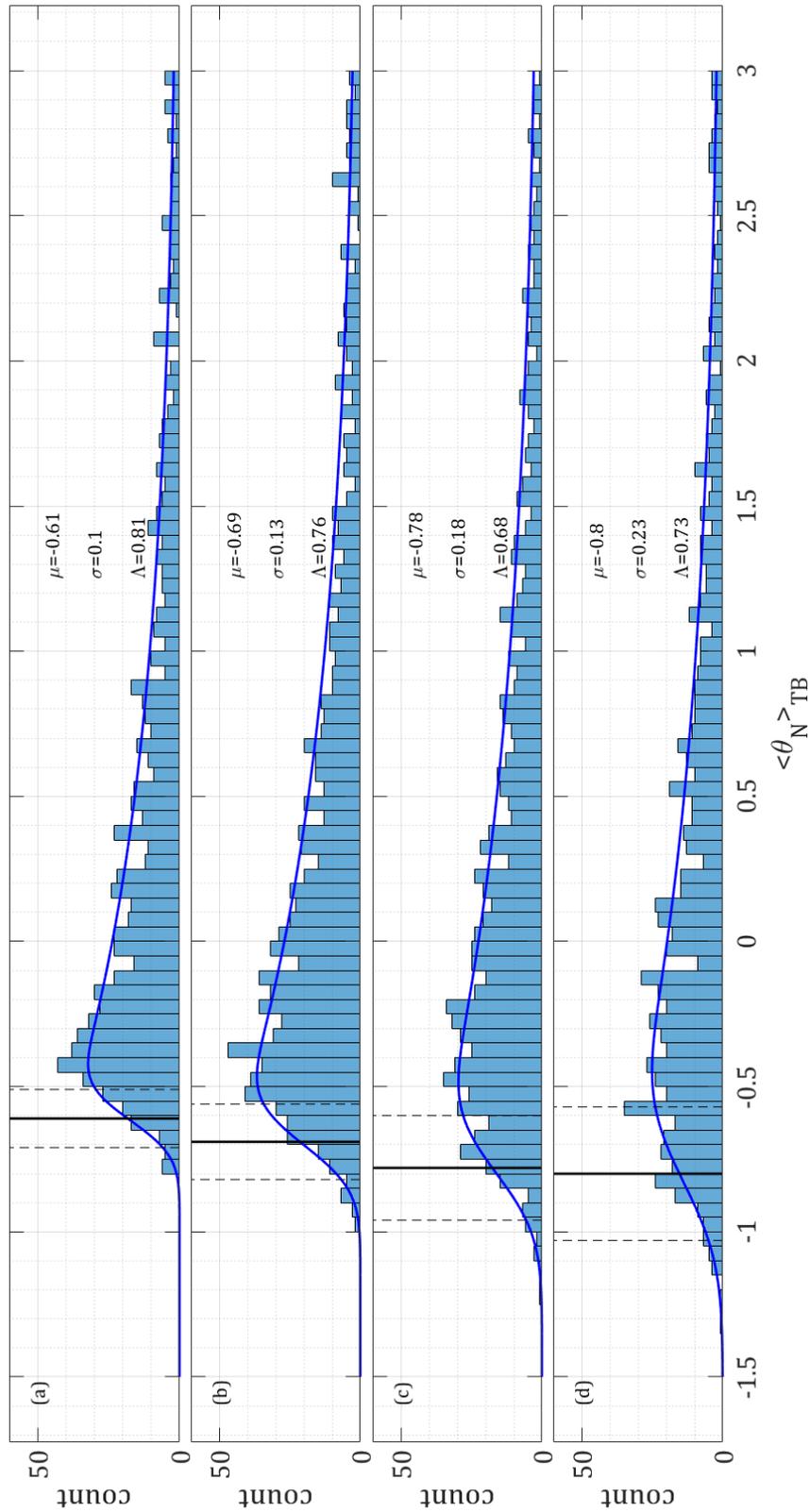

**Figure 15** *Distribution of average $\langle \theta_N \rangle_{TB}$ for each identified bursting period for a given a $\tau^*$ value. The blue curve represents the exponentially modified Gaussian fit of the distribution with the derived average $\mu$, standard deviation $\sigma$, both plotted on the distribution as the black full and dashed lines, respectively. The mean exponential component of the distributions is $\Lambda$. (a) Results for (a) $\tau^* = 0.5$; (b) $\tau^* = 1.0$; (c) $\tau^* = 2.0$; (d) $\tau^* = 3.0$.*

## 5. Conclusions

Motivated by the importance of gaining a better understanding of turbulent flows, and especially the need to resolve the smallest scales of turbulent fluctuations correctly, a new detection method allowing identification and quantification of elevated velocity fluctuation density, i.e. bursting periods, was presented. Addressing one of the most complex problems of identification of bursts in turbulent background flow, the new automated method is shown to be able to detect turbulent bursting periods in turbulent velocity field fluctuation records both accurately and automatically after proper selection of an appropriate threshold ($\varepsilon_T$) and resolution ($\tau^*$) based on the initial assessments using the velocity time series. Building upon the method of Kit *et. al.* [4], the identification of bursts is achieved by marking the time periods of elevated instantaneous TKE dissipation rate levels obtained using sliding window averaging. The non-dimensional window length ($\tau^*$) is prescribed by the turbulent flow characteristics obtained in pre-processing of the velocity field data: the Taylor microscale and the typical minimal burst period length. The ensemble is then appropriately normalized providing a discrimination (a two-fold increase of the normalized instantaneous TKE dissipation rate) between bursting and burst-free periods within the turbulent velocity fluctuation ensembles.

A step-by-step guide of the automated procedure is provided below. Two input parameters should be selected prior to automation; the first is $\tau^*$ and the second is $\varepsilon_T$.

1. Visually examine the time series of velocity fluctuations and estimate the typical bursting interval length ($\tau_{ATB}$).
2. Calculate the Taylor time scale, $\tau_\lambda$, of each ensemble. Alternatively, a different characteristic time scale can be used (Kolmogorov or Horizontal time scale), with the selection based on the nature of the examined flow.
3. Select a value of $\tau^*$ from $10(\tau_\lambda/\tau_{ATB}) < \tau^* < 2$ to obtain statistics averaged over sufficiently long periods.
4. Compute the normalized instantaneous TKE dissipation rate ($\varepsilon_N$) from Equations (3)-(8) and visually examine the variations of $\varepsilon_N$.
5. Select an appropriate threshold value $\varepsilon_T$. A good starting point would be selecting a two-fold increase from background turbulence values, $\varepsilon_T = 2$. Adjust the actual threshold value after examining the detected bursts maps as in Figures 1-3 and additional statistics and spectra as in Figures 5-11.
6. Identify turbulent bursting periods and calculate the appropriate statistics.

The developed method was tested and its implementation was demonstrated using velocity field records from a field study of turbulent, thermally driven, upslope BL flow experiencing diurnal fluctuations due to solar heating of the slope [14,44]. Sensitivity to the selected averaging non-dimensional window size $\tau^*$ was examined, and it was shown that within the selected range the actual window size is of lesser importance. The results were also shown to converge with the Kit e*t. al.* [4] results for the window size corresponding the total velocity record ensemble length. Statistical examination of the results, including exceedance probability function and spectral analysis of bursting and burst-free periods, strongly supported the selection of a two-fold threshold for the normalized instantaneous TKE variations for detection of bursting periods in the examined data. Spectra of burst-free periods showed characteristics closely resembling those of Kolmogorov's theory, while the bursting periods demonstrated significant deviations. We have presented the distribution of the Taylor microscale and the velocity derivative skewness for each period separately as empirical evidence for the change in flow behaviour in the duration of turbulent busting events. The burst-free periods were characterized by the velocity derivative skewness values resembling that of -0.4, as observed in wind tunnel studies of homogeneous freely decaying turbulence. Moreover, the Taylor microscale average values during bursting periods were observed to be approximately twice smaller than those during burst-free periods. All together these findings indicated a successful identification and distinction between bursting periods and burst-free background turbulence.

For future implementation, it is recommended to apply the above mentioned statistical analysis of window size sensitivity and the conditional spectral analysis for selection of the best suitable threshold and window size values, in view of the turbulent flow characteristics and the nature of burst generation mechanism of any examined flow field. Implementation of the new method ensued successful identification of bursting periods longer than $\tau_{min}^* = 0.5$ and allowed investigation of the temperature dissipation rate behaviour in correlation with the occurrence of bursts. A significant decrease in the temperature gradient variance was found to accompany most of the bursting periods indicating the better-mixed nature of the flow in the duration of a bursting event relative to the background turbulence.

Implementation of the averaging window size selection, based on the flow parameters and proper normalization of the TKE dissipation rate variations, render the presented bursting period identification method as potentially suitable for implementation in records across different flows, including flows characterized by significantly varying mean velocity, background turbulence intensity, and forcing conditions. Proper normalization of the TKE dissipation rate allows automated identification of bursts without the need of selection of physics-based thresholds and manual identification of suitable ensembles. Based on analysis of velocity field fluctuations obtained in a single point, this method is suitable for use with data produced in laboratory, field experiments, and that of detailed direct numerical simulations. The new method opens the possibility to produce analysis of new and existing data for the purpose of investigating possible burst generation mechanisms, instabilities, scalar transport, and more to assist in improving NWP and GCM. In this work, we have also demonstrated the possible technique for producing correlations between the scalar dissipation rate and the occurrence of bursts. A proper normalization of scalar dissipation rate was suggested for future studies to better compare with the bursting period duration.

**Acknowledgements** The authors gratefully acknowledge the support of this study by the United States–Israel Binational Science Foundation under Grant 2014075. Also, the authors wish to acknowledge the valuable and insightful comments of two anonymous reviewers, who's comprehensive and challenging questions have helped to eradicate the inaccuracies and to improve the manuscript content and readability.

## Appendix

```matlab
function [EpsilonN,ThetaN] = GettingNormCurves(u,v,w,T,Fs,Tau,TauATB,nu,alpha)
%       MATLAB® code written in version 2018b.
%u      instantaneous streamwise velocity component, m/s
%v      instantaneous longitudinal velocity component, m/s
%w      instantaneous transverse velocity component, m/s
%T      instantaneous temperature, degrees Celsius
%Fs     sampling frequency, Hz
%Tau    non-dimensional window size
%TauATB approximated bursting period length from visual observation, s
%nu     kinematic viscosity, m^2/s
%alpha  thermal diffusivity of the fluid, m^2/s

%EpsilonN   Normalized instantaneous TKE dissipation rate
%ThetaN     Normalized instantaneous temperature (scalar) variance dissipation rate

N=round(Fs*Tau*TauATB,0); %must have an integer for a window size

%% Defining normalized TKE dissipation rates
um=movmedian(u,N,'omitnan');
vm=movmedian(v,N,'omitnan');
wm=movmedian(w,N,'omitnan');
ut=u-um;
vt=v-vm;
wt=w-wm;

% Equations (4)-(6) from the text
eu=(15*nu./um(1:end-1).^2).*movmedian((diff(ut).*Fs).^2,N,'omitnan');
ev=(7.5*nu./um(1:end-1).^2).*movmedian((diff(vt).*Fs).^2,N,'omitnan');
ew=(7.5*nu./um(1:end-1).^2).*movmedian((diff(wt).*Fs).^2,N,'omitnan');
% Equation (3) from the text
epsilon_m=(eu+ev+ew)./3;
% Equation (7) from the text
epsilon_m_t=(epsilon_m-nanmedian(epsilon_m));
% Equation (8) from the text
EpsilonN=epsilon_m_t./rms(epsilon_m_t);

%% Defining normalized scalar (temperature) variance dissipation rate
Tm=movmean(T,N,'omitnan');
Tt=T-Tm;
%define the instantaneous temperature variance dissipation rate
[Tx,~]=gradient(Tt);
% Equation (10) from the text
Theta=(3.*alpha./um.^2).*movmean((Tx.*Fs).^2,Num,2,'omitnan');

% Equation (12) from the text
Thetat= Theta-nanmean(Theta,2);
% Equation (11) from the text
ThetaN= Thetat./rms(Thetat,2);

end
```